\documentclass[showpacs,oneside,twocolumn,prl,amsmath,amssymb,fleqn]{revtex4-1}
\usepackage{latexsym}
\usepackage{amsthm}
\usepackage{amssymb}
\usepackage{amsmath}
\usepackage[all]{xy}
\usepackage{graphicx}
\usepackage{subfigure}
\usepackage{dcolumn}
\usepackage{bm}
\usepackage{bbm}
\usepackage{color}
\usepackage{amsfonts}
\usepackage{cases}
\usepackage{amsfonts}
\usepackage{array}
\usepackage{booktabs}
\usepackage[a4paper,pagebackref=true,colorlinks=true,
linkcolor=blue,citecolor=blue,
pdfauthor={ },
pdftitle={ },
pdfsubject={ },
pdfkeywords={ }]{hyperref}

\begin{document}
\title{Implementation of dynamically corrected gates for single-spin qubits}
\author{Xing Rong}
\author{Jianpei Geng}
\author{Zixiang Wang}
\author{Qi Zhang}
\author{Chenyong Ju}
\author{Fazhan Shi}
\author{Chang-Kui Duan}
\altaffiliation{ckduan@ustc.edu.cn}
\author{Jiangfeng Du}
\altaffiliation{djf@ustc.edu.cn}
\affiliation{Hefei National Laboratory for Physics Sciences at Microscale and Department of Modern Physics, University of Science
and Technology of China, Hefei, 230026, China}

\begin{abstract}
Precise control of an open quantum system is critical to quantum information processing, but is challenging due to inevitable interactions between the quantum system and the environment. We demonstrated experimentally at room temperature a type of dynamically corrected gates on the nitrogen-vacancy centers in diamond.
 The infidelity of quantum gates caused by environment nuclear spin bath is reduced from being the second-order to the sixth-order of the noise to control field ratio, which offers greater efficiency in reducing the infidelity by reducing the noise level. The decay time of the coherent oscillation driven by dynamically corrected gates is shown to be two orders of magnitude longer than the dephasing time, and is essentially limited by spin-lattice relaxation. The infidelity of DCG, which is actually constrained by the decay time, reaches $4\times 10^{-3}$ at room temperature and is further reducible by 2-3 orders of magnitudes via lowering temperature. The greatly reduced noise dependence of infidelity and the uttermost extension of the coherent time mark an important step towards fault-tolerant quantum computation in realistic systems.
\end{abstract}
\maketitle

Quantum information processing can provide a dramatic speed-up over classical computer for certain problems \cite{ref1}.
One of the most urgent demands in quantum computation is to realize noise-resistant universal quantum gates for qubits.
Strategies including quantum error correction \cite{ref13,ref14,ref15}, decoherence-free subspace \cite{ref16,ref17},
and dynamical decoupling (DD) \cite{ref5} have been developed to accomplish this task.
Compared with the other two strategies, the resource requirements for the DD are modest \cite{ref18}, and no extra qubits is required. It uses stroboscopic
qubit flips to average out the coupling to the environment.
There are experiments to demonstrate DD as a successful technique in preventing the quantum states from being destroyed in open systems \cite{ref6,ref7,ref8}. However, implementing DD in quantum gates is still of challenge, because
in general the gate operation may not commute with the DD control. Recently, there are theoretical studies to address this problem
via dynamically corrected gates (DCGs) \cite{ref9,ref10,ref11,ref24}, but experimental implementation is still elusive.

 Herein, we experimentally demonstrated a type of DCG, namely SUPCODE\cite{ref11}, in a single electron spin in diamond at room temperature.
We experimentally verified that the SUPCODE reduces the error stemming from magnetic field fluctuation to the sixth-order of the fluctuation field to control field ratio, and largely diminishes the decoherence effect.
The performance of quantum gate has been protected far beyond the quantum system's dephasing time $T_2^*$ and is approximately
limited by the spin-lattice relaxation time $T_1$. Our successful demonstration of decoherece-protected universal quantum
control is important for future quantum computations based on solid-state devices.

\begin{figure}
\centering
\includegraphics[width=0.95\columnwidth]{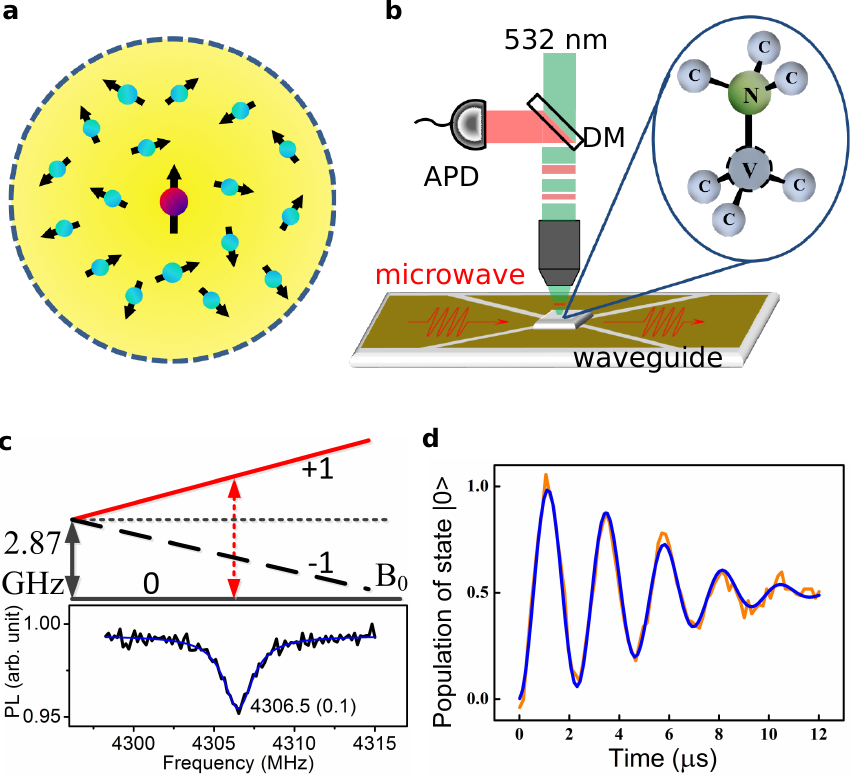}
    \caption{(Color online)
        \label{fig1}
    System for measuring the coherent dynamics of the NV center spin.
   (a) A spin qubit interacting with the nuclear spin bath.
   (b) Schematic of the ODMR setup. The external magnetic field $B_0$ was applied by a permanent magnet while the microwave pulse was carried by waveguide.
   (c) Energy level diagram of N-V center with external magnetic field $B_0$ (upper panel) and continuous-wave spectrum of $|0\rangle\leftrightarrow|1\rangle$ with $B_0 = 513~$Gauss (lower panel).
   (d) FID of the electron spin. Experimental data (orange line) are fitted by $y = 0.5- 0.5 \exp [-(t/T_2^*)^2 ] \cos(2\pi\omega t)$ (blue line) with  $\omega=0.5~$MHz and  $T_2^*=6.56 (0.17)\mu s$.
}
\end{figure}

 The Hamiltonian describing a general single-qubit unitary operation on an electron-spin qubit
in the rotating frame can be described as $H_C = \omega_1 \mathbf{n}\cdot \mathbf{S}$,
where $\mathbf{S}=(S_x,S_y,S_z)$ is the spin vector operator of the qubit, $\mathbf{n}$
is a three-dimensional vector, and the strength $\omega_1$ is a real parameter.
When the qubit is subjected to the noisy environment, the total Hamiltonian can be expressed as $H =H_S + H_{SB} + H_B + H_C$,
where $H_S = \Omega_0S_z$ is the system Hamiltonian and $\Omega_0 $ is the
off-resonance frequency. $H_{SB}$ stands for the qubit-environment coupling and $H_B$ is the environmental Hamiltonian.
Herein, we describe the environment as a nuclear spin bath and the coupling as a pure dephasing interaction
$H_{SB} = \sum_k b_kS_zI_z^k$, where $I_z^k$ is the spin operator of the $k$th nuclear spin and
the $b_k$ is the strength of the hyperfine interaction between the qubit and the $k$th nuclear spin.

Fig. \ref{fig1}a depicts an electron spin that interacts with the nuclear spin bath.
The hyperfine interaction with the nuclear spins results in a random local magnetic field (Overhauser field)
$\delta = \sum_{k} b_kI_z^k$ of typical strength the order of magnitude of $1~$MHz in solids.
The thermal distribution of the Overhauser field causes rapid free induction decay (FID) of the electron spin coherence.
There are also dynamical fluctuations of the local Overhauser field driven by pair-wise nuclear-spin flip-flop.
Because the dynamical fluctuations in $\delta $ are much slower than the typical gate time,
we take $\delta $ as a random time-independent variable\cite{ref11,ref19}.

For simplicity, we consider the on resonance case ($\Omega_0=0$) and set the axis $\mathbf{n}$ to ${\it x}$ axis. Then the Hamiltonian $H$ can be written as
$H =\delta S_z + \omega_1S_x$.
We define the fidelity of the gate
$F = \rm{Tr}(AB^{-1})/2 $, where $A= \exp (-i 2\pi\omega_1 S_x \tau) $
is the ideal gate operation and $B=\exp (-i 2\pi H \tau) $ is the gate operation in the presence of dephasing noise,
where $\tau$ is the gate time. The case of interest in implementing DCGs is $\delta\ll\omega_1$.
The infidelity of the gate, $\Delta = 1- F$, for the plain case (rectangle pulse), is of the second order
in $\delta/\omega_1$ (see Table \ref{table1}).
Since the nuclear spins flip, the qubit experiences a different effective $\delta S_z$ in every experiment once a new sequence is started.
When we average over many sequences as required to build statistics, $\delta S_z$ will cause the decay of the Rabi oscillations even in the absence of the instability of control field $H_C$\cite{ref19}.

\begin{figure}
\centering
\includegraphics[width=0.95\columnwidth]{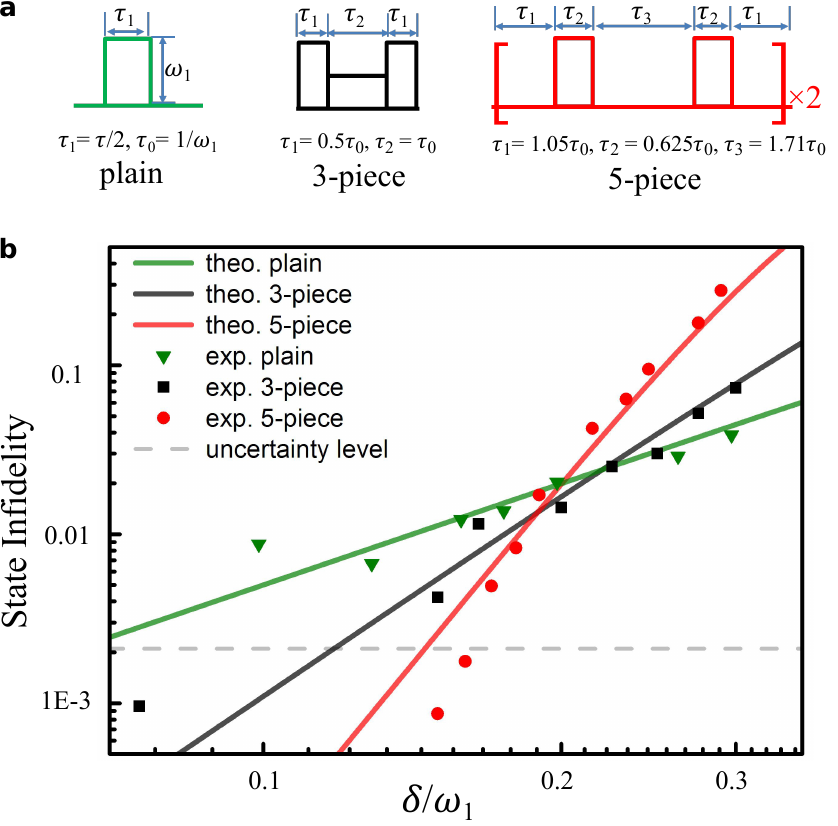}
    \caption{(Color online)
      \label{fig2}
    {\bf The performance of the SUPCDE $\pi$ gate with $\delta/\omega_1$.}
   ({\bf a}) Pulse sequences corresponding to plain sequence,
   three- and five-piece SUPCODE sequences.  The time durations $\tau_i$ ($i=1,2,3,{\rm ~and~} 4$) are calculated according to Ref. \onlinecite{ref11}
   with $\tau_0 = 1/\omega_1$ and $\omega_1 = 20~$MHz.
   ({\bf b}) The measured and theoretically predicted state infidelities of the
   destination states as functions of $\delta/\omega_1$.
   The uncertainty level of the state fidelity due to the statistical fluctuation of photon
   is plotted as a gray dashed line. Experimental data fit the theoretical predictions (lines) well.
 }
\end{figure}

SUPCODE protects the qubit from the decoherence effect aroused from the nuclear spin bath,
while at the same time provides a universal quantum control over a single qubit.
Herein, we take five-piece SUPCODE as an example.
We denote a rectangle pulse as $R_\phi(\theta)$, which stands for a $\theta$ rotation around about an in-plane axis $\phi$.
The sequence of five-piece SUPCODE $\pi$ gate is [$(\tau_1-R_0(2\pi\cdot\omega_1\tau_2)-\tau_3-R_0(2\pi\cdot\omega_1\tau_2)-\tau_1)\times 2$] , where $\tau_1 = 1.05~\tau_0 $, $\tau_2 = 0.625~\tau_0 $ and $\tau_3 = 1.71~\tau_0  $, with $\tau_0 = 1/\omega_1$.
The details of other pulse sequences are included in Supplemental Material\cite{SM}.
According to Ref. \onlinecite{ref11}, the infidelity of the five-piece SUPCODE gate is of the sixth order in $\delta/\omega_1$.
Table \ref{table1} summarized the infidelity of several SUPCODE $\pi$ gates, whose pulse sequences are plotted in Fig. \ref{fig2}a.
Due to a large coefficient in the leading term, nine-piece SUPCODE only provides better performance when $\delta/\omega_1 \lesssim 1\%$ (see \cite{SM}),
where the infidelity of five-piece SUPCODE is already below $10^{-4}$,
the threshold for fault-tolerant quantum computation\cite{ref1}.
Hence we focus on five-piece SUPCODE, whose results are plotted in Fig. \ref{fig2}b.
It is noted that SUPCODE also reduces the complexity of the control pulse, e.g.,
five-piece SUPCODE sequence requires only two rectangle pulses with finite amplitudes
and three waiting time durations, without phase switching during the gate time.
Thus it makes SUPCODE a practical dynamically corrected gate in solid-state systems in realistic scenarios.

\begin{table}
\caption{Infidelity of SUPCODE $\pi$ gate as function of $\delta/\omega_1$. The left column shows the types of the pulse sequence, where 'plain' stands for normal rectangle pulse and '$n$-piece' is short for the $n$-piece SUPCODE pulse.}
\begin{ruledtabular}
\begin{tabular}{cc}
  \hspace{2em} Sequence \hspace{1em} & \hspace{6em}  Infidelity $\Delta$   \hspace{6em}    \\
  \hline
  plain		&     \hspace{1.5em}$0.5(\delta/\omega_1)^2$ + O$(\delta/\omega_1)^{4}$   \\
  three-piece		&     \hspace{1em}$11.1(\delta/\omega_1)^4$ + O$(\delta/\omega_1)^6$   \\
  five-piece          	&     \hspace{1em}$64.1(\delta/\omega_1)^6$ + O$(\delta/\omega_1)^{8}$\\
  nine-piece          	&     \hspace{0.2em}$317237(\delta/\omega_1)^8$ + O$(\delta/\omega_1)^{10}$ 
  \label{table1}
\end{tabular}
\end{ruledtabular}
\end{table}

The setup and experimental scheme are schematically shown in Fig. \ref{fig1}b.
A $B_0$ of $513~$Gauss was adopted so as to achieve effective polarization of the nitrogen nuclear spin in NV center,
which is confirmed by the continuous wave spectrum shown in Fig. \ref{fig1}c.
Quantum states $|0\rangle$ and $|1\rangle$ are encoded as a qubit, which can be manipulated
via microwave pulses with frequency $4306.5~$MHz.
The undesired couplings between the qubit and the surrounding $^{13}C$ nuclear spins lead to the dephasing effect and therefore reduces the fidelity of the quantum operations.
The resulted FID of the qubit is depicted in Fig. \ref{fig1}d, where a dephasing time $T_2^*$ = 6.56 (0.17) $\mu$s is obtained.

We first scrutinize the performance of SUPCODE under dephasing noises.
We experimentally investigated the $\pi$ rotation about the ${\it x}$ axis (NOT gate) via plain pulse, three- and five-piece SUPCODEs.
The state of the electron spin was first initialized to $|\Psi_0\rangle = |0\rangle$ by laser.
After applying the NOT gate on the initial state, the state fidelity\cite{ref1} of the resultant state $\rho_f$ is $f= \sqrt{\langle 1|\rho_f|1\rangle}$ by definition and
the infidelity of state is $\delta f = 1-f$.  We are aiming to investigate the robustness of SUPCODE
against the noise stemming from the quasi static fluctuation of the magnetic field, which is simulated
by detuning the frequency of the microwave from the on resonance frequency.
The value of $\delta$ ranges from $0$ to $3~$MHz and the strength of the control
field is set at $20~$MHz, which is verified by Rabi nutation experiments.
The three different pulse sequences for the $\pi$ gate are given in
Fig. \ref{fig2}a and their performance are plotted as functions of $\delta/\omega_1$ in Fig. \ref{fig2}b.
For the three-piece SUPCODE pulse, the strength of the second pulse is $\omega_1/2$.
For the case of five-piece SUPCODE, the $\pi$ rotation is realized by two five-piece SUPCODE $\pi/2$ gates.
Lines are theoretical predictions, while green triangles, black rectangles and red circles are the experimental data for plain pulse,
three- and five-piece SUPCODE $\pi$ pulses, respectively.
Each data point is averaged by $7.2\times 10^7$ times and the uncertainty due to
the statistical fluctuation of photon number has been plotted as dashed line.
The experimental data fit the theoretical predictions well. For the case of
NOT gate on the initial state $|0\rangle$, the infidelity of the states $\delta f$ equals to the infidelity of quantum gates $\Delta$ (see \cite{SM}).
Our results show that three-piece (five-piece) SUPCODE pulses reduce the error to the fourth (sixth) order in $\delta/\omega_1$, in agreement with theoretical predictions\cite{ref11}.

\begin{figure}
\centering
\includegraphics[width=0.95\columnwidth]{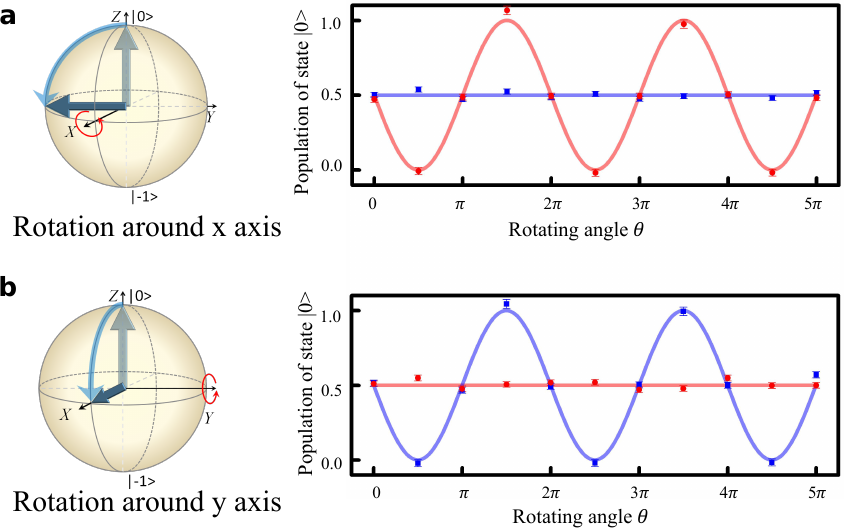}
    \caption{(Color online)
    Universal control of a single spin by five-piece SUPCODE pulse.
   The Bloch-sphere diagrams of the spin vectors rotating around
   $\mathbf{x}$ (a) and $\mathbf{y}$ (b), and the detected coherent oscillations of the
    qubit driven by the corresponding five-piece SUPCODE pulses.
    Light red (blue) lines are the theoretical predictions with X (Y) detection and
    red (blue) rectangles are the corresponding experimental data.
    \label{fig3}
 }
\end{figure}

Secondly, we show that universal single-qubit gates can be achieved by SUPCODE.
Universal control of a single qubit, which is of vital importance for quantum computation, requires the ability to realize precisely rotations around two different axes of the Bloch sphere.
By successively applying the five-piece SUPCODE pulses on the electron spin,
we can force the spin vector to rotate along the $\it x$ ($\it y$) axis by setting the phase of the microwave pulse to $0$ ($\pi/2$).
The followed readout procedure consists of a detecting $\pi/2$ microwave pulse and a laser pulse.
The phase of the detecting microwave is set to $0$ ($\pi/2$) so that the detection is sensitive to the Y (X) coherence
of the quantum states. This readout procedure is analogous to quadrature detection in NMR. If the qubit is driven
to rotate along the ${\it x}$ (${\it y}$) axis, the spin vector will rotate in the $y$-$z$ ($x$-$z$) plane (see Fig. \ref{fig3}), and coherent oscillations are observable by setting the phase of detecting microwave pulse to $0$ ($\pi/2$),
while there will be no oscillations with phase of detecting microwave pulse being set to $\pi/2$ ($0$).
Fig. \ref{fig3} a and b plot the coherent oscillations driven by five-piece SUPCODE pulse around ${\it x}$ (${\it y}$) axis.
Red circles (blue rectangles) with error bars are experimental data under X (Y) detection and light red (light blue)
lines are from the theoretical predictions.

\begin{figure}
\centering
\includegraphics[width=0.95\columnwidth]{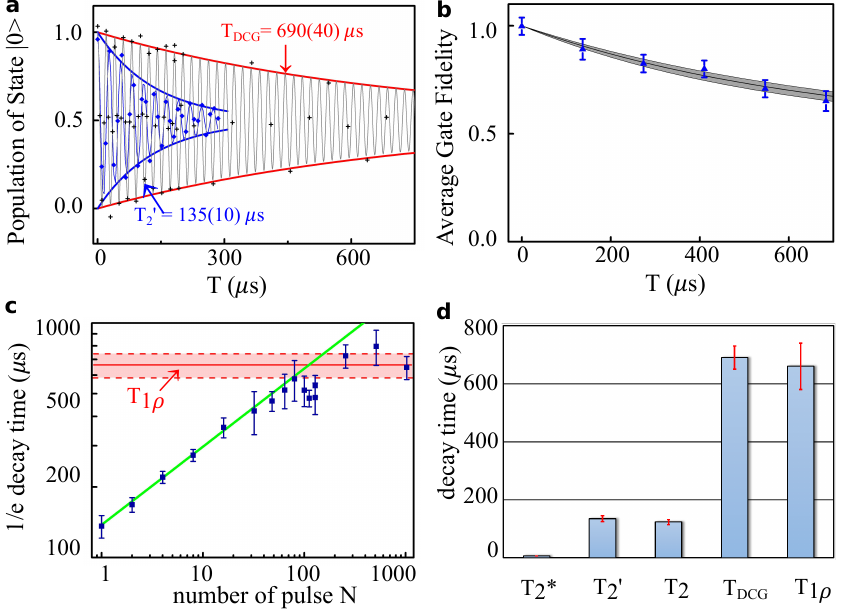}
    \caption{(Color online)
   Precise quantum control of a single electron-spin qubit towards $T_1$ limit.
   (a) Comparison of the quantum oscillation driven by five-piece SUPCODE pulses with the one by plain pulses. Black crosses are the SUPCODE experimental data and red
   lines are the envelopes of the function
   $0.5 + 0.5\exp(-T/T_{\rm DCG})\cos(\omega T)$ (gray line), with $T_{\rm DCG}= 690(40)~\mu$s. The measured (blue diamond) quantum oscillation by the plain
   pulses (Rabi oscillation) was fitted by $0.5 + 0.5\exp(-T/T_2')\cos(\omega' T)$ (blue line) with $T_2'=135(10)~\mu$s.
   (b) The decay of the fidelity of quantum gates with five-piece SUPCODE obtain via quantum process tomography. Black line is theoretically predicted fidelity when only $T_{1\rho}$ process is taken into account. The shaded region is calculated using the uncertainty of the measured $T_{1\rho}$. Blue triangles are experimental fidelities by QPT, which shows a $T_{1\rho}$ limited decay.
   (c) Extended coherence times via CPMGs. $N$ stands for the number of the $\pi$ pulses.
   After applying more than $1000$ pulses, the prolonged coherence times are limited by the $T_{1\rho} = 660(80)~\mu$s measured by spin-locking experiment.
   (d) Comparison of the decay times.
   $T_{\rm DCG}$ is not only far beyond the dephasing time $T_2^* = 6.56(0.17)~\mu$s,
   but also much longer than $T_2 = 123.2(8.8)~\mu$s determined by Hahn echo experiment, and equals $T_{1\rho}$ within the uncertainty.
    \label{fig4}}
\end{figure}

Thirdly, we experimentally show that coherence time can be greatly extended via SUPCODE. We applied the five-piece SUPCODE $\pi/2$ gate successively on the electron spin qubit.
To minimize the influence of the instability of the control field $H_C$\cite{ref19},
we adopted a microwave pulse of an amplitude corresponding to $\omega_1 = 1~$MHz.
The quantum oscillation driven by SUPCODE is plotted in Fig.~\ref{fig4}a.
A decay time constant $T_{{\rm DCG}} = 690 (40)~\mu$s is derived from the experimental data (black crosses), which is two orders of magnitude longer than the dephasing time $T_2^*$.
The Rabi oscillation (blue diamonds) driven by the normal rectangle pulse in
Fig.~\ref{fig4}a of $\omega_1 = 1~$MHz shows a decay time $T_2' = 135(10)~\mu$s, which is also much shorter  than $T_{{\rm DCG}} $.
This verified that SUPCODE can largely
suppress the dephasing effect during the gate time.

It is noted that $T_{{\rm DCG}} $ achieved here has reached the spin-lattice relaxation time in the rotating frame $T_{1\rho} =  660(80)~\mu$s, which is determined by a spin-locking experiment. $T_{1\rho}$ is the upper bound achievable by any multiple dynamical decoupling pulse sequence  for ${\it T}_2$ \cite{ref21}.
To confirm this, we applied Carr-Purcell-Meiboom-Gill (CPMG) sequences on the qubit to prolong the coherence time with $\omega_1 = 10~$MHz.
The scaling of the extended coherence times with the number of pulses $N$ is depicted in Fig. \ref{fig4}c as blue rectangles.
We found that the coherence time can be efficiently prolonged by CPMG sequences with the scaling $T_2\propto N^{0.33}$
until it approaches the saturation value for $N>100$ at $T_{1\rho} $.

Finally, we examine the performance of the quantum gate via SUPCODE  in detail by quantum process tomography (QPT).
Any quantum operation can be considered as mapping the input quantum state $\rho_{i}$
to the output $\rho_o = \varepsilon(\rho_i) = \sum\limits_{m,n} \chi_{mn}A_m\rho_i A^\dag_n$
with $\{A_m\}=\{I,\sigma_x,\sigma_y,\sigma_z\}$, where $\sigma_x,\sigma_y$ and $\sigma_z$
are Pauli operators. The matrix $\chi$ completely and uniquely describes the process $\varepsilon$
and can be experimentally reconstructed by QPT. The average fidelity of the quantum gate\cite{ref20} is defined by
\begin{equation}
F(\varepsilon, U)= \frac{1}{2} + \frac{1}{12}\sum\limits_{j = x,y,z} \rm{Tr}[U\sigma_jU^\dag\varepsilon(\sigma_j)],
\end{equation} where $U$ is the ideal operation. We applied five-piece SUPCODE $\pi/2$ gates
successively on the electron spins and performed QPT when the gates were repeated for $M =0, 27, 54, 81, 108$ and $135$ times.
The variation of gate fidelities is depicted as a function of the operation time in Fig. \ref{fig4}b.
Experimental data of the gate fidelity (blue triangles) were obtained by using QPT. We also plot the theoretical predicted variation of gate fidelities (black lines and shaded region) with only $T_{1\rho}$ relaxation being considered. The fact that the experimental data match the theoretical prediction indicates that the performance of the SUPCODE is limited by $T_{1\rho}$. The details of experimental QPT, theoretical calculations and $T_{1\rho}$ measurement can be found in Supplemental Material\cite{SM}.

$\it{Discussion}.-$ We have experimentally demonstrated dynamically corrected gates for a single electron spin in diamond at room temperature.  The related decay times are summarized in Fig. \ref{fig4}d.
The coherence time without any dynamical decoupling control pulse is $T_2^* = 6.56(0.17)~\mu$s.
It can be prolonged to $T_2 = 123.2(8.8)~\mu$s by a single refocus $\pi$ pulse.
This dynamical decoupling pulse only enables robust quantum state storage against
the dephasing effect. The Rabi oscillation driven by normal pulse, which can be used to realize quantum gate, has a decay time $T_2' = 135(10)~\mu$s when $\omega_1 = 1~$MHz.
In comparison, the decay time of the quantum oscillation via SUPCODE is $T_{\rm{DCG}} = 690 (40)~\mu$s, which equals $T_{1\rho} = 660(80)~\mu$s within the uncertainty. As $T_{1\rho}$ or the CPMG limit of $T_{2}$ can be prolonged by 2-3 order of magnitudes in proportion to $T_1$ (i.e., $\sim 0.5T_1$ \cite{Lowtemp_NV2}) by lowering the temperature in a wide range \cite{ref23,Lowtemp_NV1,Lowtemp_NV2}, the DCGs demonstrated here can be considered as approximately $T_{1}$ limited, and further improvement for the coherent time and fidelity of DCGs are feasible at a wide range of lower temperature along with the prolonging of $T_1$.


 The fidelity of each five-piece SUPCODE $\pi/2$ gate can be derived  to be very close to unity, while the gate time ($5.063~\mu$s) is comparable to $T_2^*$. Since we applied SUPCODE gates successively on the electron spin qubit, the error of each gate accumulated. The decay of the fidelity (see Fig.\ref{fig4}b) can be used to estimate each gate's fidelity. 
 We estimate the average fidelity of SUPCODE $\pi/2$ gate to be $0.9961(2)$ (see \cite{SM}), which is close to the threshold of the fault-tolerant quantum computation\cite{nature2005}. Since $T_2$ can be dramatically enhance by lowering the temperatures\cite{Lowtemp_NV2}, much longer $T_{1\rho}$ is expected at cryogenic temperatures and the fidelity of DCG will be further improved to meet the requirement for fault-tolerant quantum computation.

Our experimental implementation of DCGs is expected to have various applications in quantum information processing, high resolution spectroscopy\cite{ref2,ref3}, and various quantum metrologies\cite{ref4}, where high-fidelity quantum gates are required. The DCGs implemented here can also be applied in other important physical systems, such as singlet-triplet spin qubit in a semiconductor double dot\cite{ref11,quantdot}, phosphorus doped in silicon\cite{Kane98,PSi1,PSi2} and superconducting qubits\cite{Devoret2013,Bylander2011}, so lay the foundation for precise control of solid-state quantum systems.
Although our current work is focused on an experimental implementation of single-qubit DCGs, theoretical schemes
for two-qubit DCGs are available \cite{ref24} for experimental implementation in the future.
Our experimental implementation high-fidelity and $T_1$-limited DCGs marks an important step towards realistic fault-tolerant quantum computation.

We thank Yiqun Wang in Suzhou Institute of Nano-Tech and Nano-Bionics for fabricating the coplanar waveguide, and Ren-bao Liu from the Chinese University of Hong Kong for helpful discussions. This work was supported by the National Key Basic Research Program of China (Grant No. 2013CB921800), the National Natural Science Foundation of China (Grants Nos. 11227901, 11275183, 11104262, 91021005 and 10834005), the `Strategic Priority Research Program (B)' of the CAS (Grant No. XDB01030400) and the Fundamental Research Funds for the Central Universities.


\clearpage

\end{document}